\documentclass[12pt,onecolumn]{emulateapj}
\usepackage{amsfonts,amsmath,graphicx,natbib}
\usepackage{epstopdf}

\shortauthors{Bower et al.}
\shorttitle{Proper Motion of the GC Pulsar}
\begin{document}

\newcommand\degd{\ifmmode^{\circ}\!\!\!.\,\else$^{\circ}\!\!\!.\,$\fi}
\newcommand{\etal}{{\it et al.\ }}
\newcommand{\uv}{(u,v)}
\newcommand{\rdm}{{\rm\ rad\ m^{-2}}}
\newcommand{\msuny}{{\rm\ M_{\sun}\ y^{-1}}}
\newcommand{\mylesssim}{\stackrel{\scriptstyle <}{\scriptstyle \sim}}
\newcommand{\lsim}{\stackrel{\scriptstyle <}{\scriptstyle \sim}}
\newcommand{\gsim}{\stackrel{\scriptstyle >}{\scriptstyle \sim}}
\newcommand{\sci}{Science}
\newcommand{\nar}{New A Rev.}
\newcommand{\sgr}{PSR J1745-2900}
\newcommand{\kms}{\ensuremath{{\rm km\,s}^{-1}}}
\newcommand{\masy}{\ensuremath{{\rm mas\,yr}^{-1}}}

\def\kbar{{\mathchar'26\mkern-9mu k}}
\def\totd{{\mathrm{d}}}


\title{The Proper Motion of the Galactic Center Pulsar Relative to Sagittarius A*}

\author{
 Geoffrey C.\ Bower\altaffilmark{1}, 
Adam Deller\altaffilmark{2}, 
Paul Demorest\altaffilmark{3},
Andreas Brunthaler\altaffilmark{4}, 
Heino Falcke\altaffilmark{5,2,4},
Monika Moscibrodzka\altaffilmark{5},
Ryan M. O'Leary\altaffilmark{6},
Ralph P. Eatough\altaffilmark{4},
Michael Kramer\altaffilmark{4,7},
K.J. Lee\altaffilmark{4},
Laura Spitler\altaffilmark{4},
Gregory Desvignes\altaffilmark{4},
Anthony P. Rushton\altaffilmark{8,9},
Sheperd Doeleman\altaffilmark{10,11},
Mark J. Reid\altaffilmark{11}
}

\altaffiltext{1}{Academica Sinica Institute of Astronomy and Astrophysics, 645 N. A'ohoku Place,Hilo, HI 96720, USA; gbower@asiaa.sinica.edu.tw}
\altaffiltext{2}{ASTRON, P.O. Box 2, 7990 AA Dwingeloo, The Netherlands}
\altaffiltext{3}{NRAO, 520 Edgemont Road, Charlottesville, VA 22903-2475, USA}
\altaffiltext{4}{Max-Planck-Institut f\"ur Radioastronomie, Auf dem H\"ugel 69, D-53121 Bonn, Germany}
\altaffiltext{5}{Department of Astrophysics, Institute for Mathematics, Astrophysics and Particle Physics (IMAPP), Radboud University, PO Box 9010, 6500 GL Nijmegen, The Netherlands} 
\altaffiltext{6}{JILA, University of Colorado and NIST, 440 UCB, Boulder, CO 80309-0440, USA}
\altaffiltext{7}{Jodrell Bank Centre of Astrophysics, University of Manchester, Manchester M13 9PL, UK}
\altaffiltext{8}{Department of Physics, Astrophysics, University of Oxford, Keble Road, Oxford OX1 3RH, UK}
\altaffiltext{9}{School of Physics and Astronomy, University of Southampton, Highfield, Southampton SO17 1BJ, UK}
\altaffiltext{10}{MIT Haystack Observatory, Route 40, Westford, MA 01886, USA}
\altaffiltext{11}{Harvard-Smithsonian Center for Astrophysics, 60 Garden Street, Cambridge, MA 02138, USA}

\begin{abstract}

We measure the proper motion of the pulsar \sgr\ relative to the Galactic Center massive black hole, Sgr A*, using the Very Long Baseline Array (VLBA).
The pulsar has a transverse velocity of $236 \pm 11\, \kms$ at position angle $22 \pm 2$ deg East of North at a projected
separation of 0.097 pc
from Sgr A*.  
Given the unknown radial velocity, this transverse velocity measurement does not conclusively prove that the pulsar is bound to Sgr A*; however, the probability of chance alignment is very small. 
We do show that the velocity and position is consistent with a bound orbit originating in the clockwise disk of massive stars 
orbiting Sgr A* and a natal velocity 
kick of $\lsim 500\, \kms$. An origin among the isotropic stellar cluster is possible but less probable. 
If the pulsar remains radio-bright,
multi-year astrometry of \sgr\ can detect its acceleration and determine the full three-dimensional orbit. 
We also demonstrate that \sgr\ exhibits the same angular broadening as Sgr A* over a wavelength range of 3.6 cm to 0.7 cm, further
confirming that the two sources share the same interstellar scattering properties.
Finally, we place the first 
limits 
on the presence of a wavelength-dependent shift in the position of Sgr A*, i.e., the core shift, one of the expected
properties of optically-thick jet emission.  Our results for \sgr\ support the hypothesis that Galactic Center pulsars will originate
from the stellar disk and deepens the mystery regarding the small number of detected Galactic Center pulsars.
\end{abstract}


\section{Introduction}
\label{sec:intro}

The discovery of the pulsar \sgr\ at a projected separation of 0.1 pc from the massive black hole in the Galactic
Center, Sgr A*, provides an unprecedented opportunity to explore stellar evolution, the population of compact objects,
and the interstellar medium of the Galactic Center.  Ultimately, these elements contribute to our understanding
of the possibility for the use of pulsars in short-period orbits to characterize the space-time metric of the black 
hole \citep[e.g.,][]{1986ARA&A..24..537B,1996A&A...311..746W,2004ApJ...615..253P,2004NewAR..48.1413C,2012ApJ...747....1L}.  
In addition, the proximity of the pulsar to Sgr A* provides a unique reference source to characterize the
high angular resolution properties of Sgr A* that are obscured by the effects of interstellar scattering 
\citep{1992ApJ...396..686V,1994ApJ...427L..43F,1998ApJ...508L..61L,2004Sci...304..704,2006ApJ...648L.127B,2014ApJ...790....1B}.

\sgr\ was discovered serendipitously on 24 April 2013 (MJD$=56406$) through detection of a strong X-ray burst
as part of a daily monitoring campaign carried out by the {\em Swift} satellite 
\citep{2013ApJ...770L..24K}.  NuSTAR observations then detected periodic flux variations
with $P=3.76$ s and a hydrogen absorption column $N_H \sim 10^{23}$ cm$^{-2}$ that is
characteristic of a location in the Galactic Center 
\citep{2013ApJ...770L..23M}. {\em Chandra} observations localized the source 
offset from Sgr A* by $\sim 3$ arcseconds, a projected separation
of 0.1 pc \citep{2013ApJ...775L..34R}.  \sgr\ increased in luminosity by a factor of $\gsim 10^3$ over 
upper limits from deep {\em Chandra} observations \citep{2009ApJS..181..110M}.  The measured period
derivative implies a characteristic age of $9$ kyr and a magnetic field $\sim 10^{14}$ G,
under standard assumptions for magnetic dipole breaking.  The transient nature
of the X-ray flux and the low spin-down power relative to the X-ray luminosity 
indicate that \sgr\ is not a rotation-powered pulsar but a magnetar.
Recently,
\citet{2014ApJ...786...84K} demonstrated a factor of $\sim 3$ increase in the spin-down rate of the magnetar with
hard X-ray observations.  The variability of the spin-down rate demonstrates that the characteristic age is accurate
to at most an order of magnitude.  Association of soft gamma-ray repeaters (SGRs) and anomalous X-ray pulsars (AXPs)
with supernova remnants suggests that magnetar ages are $\lsim 10^4$ -- $10^5$ yr \citep{2002ApJ...574..332T}.  
The instability of the magnetar spin limits the degree to which 
general relativistic effects can be studied through timing observations.

Pulsed radio emission from \sgr\ was discovered shortly after the pulsed X-ray emission \citep{2013Natur.501..391E,2013MNRAS.435L..29S}.
\sgr\ was found to have the largest dispersion measure, $DM=1778 \pm 3 {\rm\ pc\ cm^{-3}}$, and rotation measure,
$RM=-66960 \pm 50 {\rm\ rad\ m^{-2}}$, of any known pulsar.  These large values are consistent with the source being
embedded in the dense, magnetized plasma of the Galactic Center.  High resolution very long baseline interferometry (VLBI)
demonstrated that the pulsar also shares the extreme angular broadening 
of Sgr A* \citep{2014ApJ...780L...2B}.  
Pulsar timing show that the pulse width scales as $\nu^4$ and that pulses can be detected to frequencies below 2 GHz
\citep{2014ApJ...780L...3S}.  
Three other pulsars in the central 40 pc of the Galaxy show temporal broadening
that is comparable to that of \sgr\, i.e., $\sim 1$ sec at 1 GHz \citep{2006MNRAS.373L...6J,2009ApJ...702L.177D}.
Using the model of a geometrically thin scattering screen in combination with angular and temporal broadening,
\citet{2014ApJ...780L...2B}  demonstrate that a substantial fraction
of the scattering must occur at distances of kiloparsecs away from the Galactic Center; 
alternate explanations that lead to large temporal scattering may still be found
\citep{2014ApJ...780L...3S} 

If the scattering screen is in fact distant from the Galactic Center as the most 
straight-forward interpretation suggests,
then the shared scattering properties of \sgr\ and Sgr A*, and the detection of radio
pulses at frequencies as low as 1.6 GHz indicate that pulsars bound to Sgr A* and,
likely, throughout the Galactic Center can be readily detected.  
Strikingly, such pulsars have not been found in targeted radio searches over
a wide range of frequencies
\citep[e.g.,][]{2000ASPC..202...37K,2006MNRAS.373L...6J,2009ApJ...702L.177D,2010ApJ...715..939M,2013IAUS..291...57S,2013IAUS..291..382E}.  
Previously, the absence of Galactic Center pulsars was attributed to extreme temporal broadening at low
radio frequencies and the steep spectra of radio pulsars.  
Theory, however, has predicted populations hundreds to thousands of pulsars bound to Sgr A* based on the high rate of star formation
and the large massive star population \citep[e.g.,][]{2004ApJ...615..253P,2006ApJ...649...91F,2012ApJ...753..108W,2014ApJ...784..106Z}. 
The massive stars may have produced $\sim 20$ pulsar wind nebulae in the central 20 pc, as seen through
extended X-ray emission \citep{2008ApJ...673..251M}.
Finally, while pulsars have not been discovered, a dense cluster of compact 
objects within the central parsec, many known to be black hole binaries, has been detected 
through transient X-ray and radio emission \citep{2005ApJ...622L.113M,2005ApJ...633..218B}.

The absence of pulsar discovery has raised the issue 
of whether there is a problem of missing pulsars in the Galactic Center.  
This has led to discussion of alternative pulsar populations in the Galactic Center.
\citet{2014ApJ...783L...7D} argue that the detection of a rare radio magnetar and the demonstration of
the ability to detect ordinary pulsars implies that the pulsar population of the Galactic Center may be
distinct from the field population.  
\citet{2014arXiv1405.1031B} argue that dark matter could interact with neutron stars and lead them to
collapse into black holes.
Ultimately, deeper surveys will be required to determine whether the population is truly peculiar \citep{2014MNRAS.440L..86C}.

In this paper, we measure the astrometric properties of the pulsar \sgr.  The proper motion 
and acceleration provide important constraints on the origin of the pulsar.  In particular, we 
explore whether the pulsar has originated in the Galactic Center cluster stellar disk, which theory
has predicted as the origin for the bulk of the pulsar population.  In addition, the proper motion
can be used to translate time-domain variations in the propagation quantities (DM, RM, angular broadening,
temporal broadening) into linear
units.  This will provide a powerful probe of the length-scale of turbulent properties in the Galactic Center.

Simultaneously, we use astrometric measurements of \sgr\ relative to Sgr A* to constrain the 
radiative properties of Sgr A*.  Sgr A* shows some of the most extreme scattering
properties of any radio source with an angular size of $\sim 0.5$ arcsec at 21 cm and scaling with
$\lambda^2$ \citep{2004Sci...304..704,2006ApJ...648L.127B,2014ApJ...790....1B}.  
The angular broadening obscures the underlying physical processes such
that we are not able to determine whether a jet or accretion disk is responsible for the 
nonthermal emission.  Our new observations provide the most accurate measurements of 
the size of Sgr A* at wavelengths of 2.0 and 3.5 cm.
In addition, jet theory predicts that the core of the radio emission will shift
as a function of wavelength due to differences in the opacity \citep{1979ApJ...232...34B,1999ASPC..186..148F}.  
This ``core shift'' has been seen
in other radio sources powered by jets, most notably M81 and M87 \citep{2004ApJ...615..173B,2011Natur.477..185H}.
Our phased-reference measurement of the position of \sgr, which is intrinsically
point-like with a position independent of wavelength, therefore, provides a unique 
probe of the wavelength-dependent structure of the reference source, Sgr A*. 
These observations are complementary to short wavelength VLBI observations 
\citep[$\lambda\gsim 1.3$mm;][]{2008Natur.455...78D,2011ApJ...727L..36F} which will image 
the inner accretion disk and/or the base of the jet, and explore smaller-scale frequency-dependent
general relativistic effects \citep{2006ApJ...636L.109B}.  

In Section~\ref{sec:obs} we present our Very Long Baseline Array (VLBA) observations spanning the
first year since the discovery of the pulsar.  In Section~\ref{sec:astrometry}, we discuss results
for the motion and origin of the pulsar and conclude that it is likely that the pulsar originated in the
stellar disk.  In Section~\ref{sec:sgra}, we present results on the angular broadening and core shift.
We provide our conclusions in Section~\ref{sec:conclusions}.

\section{Observations and Data Reduction}
\label{sec:obs}

\begin{deluxetable}{cllllllllll}
\tablecaption{Observations of \sgr}
\tablewidth{0pt}
\tablehead{
\colhead{Epoch} & \colhead{Observing} 	& \colhead{Observing frequency} & \colhead{BR} & \colhead{FD} & \colhead{KP} & \colhead{LA} & \colhead{NL} & \colhead{OV} & \colhead{PT} & \colhead{Y\tablenotemark{A}} \\
\colhead{(MJD)} & \colhead{band} 		&\colhead{(GHz)} 
}
\startdata
56422 & X		& 8.540 -- 8.796 		& $\surd$ &$\surd$ &$\surd$ &$\surd$ &$\surd$ &$\surd$ &$\surd$ & D \\
56444 & X		& 8.540 -- 8.796 		& $\surd$ &$\surd$ &$\surd$\tablenotemark{B} &$\surd$\tablenotemark{B} &$\surd$ &$\surd$ &$\surd$ & C\tablenotemark{C} \\
56473 & Ku	& 15.240 -- 15.496 			& $\surd$ & \dots  &$\surd$ &$\surd$ &$\surd$ &$\surd$ &$\surd$ & C \\
56486 & X		& 8.540 -- 8.796 		& $\surd$ & \dots  &$\surd$ &$\surd$ &$\surd$ &$\surd$ &$\surd$ & C \\
56556 & X, Ku	& 8.540 -- 8.796, 15.240 -- 15.496 	& $\surd$ & \dots  & \dots  &$\surd$ &$\surd$ &$\surd$ &$\surd$ & B \\
56658 & X, Ku	& 8.540 -- 8.796, 15.240 -- 15.496 	& $\surd$ &$\surd$ & \dots  &$\surd$ &$\surd$ &$\surd$ &$\surd$ & BnA \\
56710 & X, Ku	& 8.540 -- 8.796, 15.240 -- 15.496 	& $\surd$ &$\surd$ & \dots  &$\surd$ &$\surd$ &$\surd$ &$\surd$ & A \\
56750 & K, Q,	& 21.792 -- 22.048, 42.768 -- 43.024 	& $\surd$ &$\surd$ &$\surd$ &$\surd$ & \dots  &$\surd$ &$\surd$ & \dots \\
56772 & X, Ku	& 8.540 -- 8.796, 15.240 -- 15.496 	& $\surd$ & \dots  &$\surd$ &$\surd$ & \dots  &$\surd$ &$\surd$ & A \\
56892 & Ku, Q   & 15.240 -- 15.496, 43.168 -- 43.424 	& $\surd$ & $\surd$&$\surd$ &$\surd$ &$\surd$ &$\surd$ &$\surd$ & D\tablenotemark{D} \\
56899 & K, Q,	& 21.792 -- 22.048, 42.768 -- 43.024 	& $\surd$ &$\surd$ &$\surd$ &$\surd$ &$\surd$ &$\surd$ &$\surd$ & \dots \\
\enddata
\tablenotetext{A}{Denotes the VLA configuration for the phased-array observation.}
\tablenotetext{B}{Disk pack problems resulted in the loss of 25\% of the data from this station.  }
\tablenotetext{C}{Problems with the array phasing resulted in the loss of 35\% of the VLA data from this epoch.}
\tablenotetext{D}{Atmospheric turbulence at the VLA caused array phasing at Q band to fail.}
\label{tab:obs}
\end{deluxetable}

The observing setup was already described in detail in \citet{2014ApJ...780L...2B}; we briefly summarize it here.  Observations were 6 hours in duration, taken under the VLBA project codes BB336, BB337 and BB339.  
Pulsar gating was used with a pulse width that varied between observations due to the changing magnetar pulse profile.  
A gate stretching from the $\sim$10\% point at the rising edge of the pulse to the $\sim$10\% point at the trailing edge of the pulse was applied, with a total duration of 100-300 ms.  At 8 GHz and above, the scattering time is far shorter than the intrinsic pulse width \citep[$<=1$ ms;][]{2014ApJ...780L...3S}, so virtually all of the scattered emission is captured.
After the observations already presented in \citet{2014ApJ...780L...2B}, we began observing at both 
Ku band (15.3 GHz) and X band (8.6 GHz) within a single observation, changing frequencies on a timescale of 8 minutes.  In addition, one 
observation of Sgr A*  at higher frequencies (22 GHz and 43 GHz; project code BR187) without the VLA was included.
For the observations of MJD 56556 onwards, we used Sgr A* as the reference source for phasing the VLA, eliminating the need to slew to an external
calibrator.  One epoch (MJD 56556) failed due to the absence of two of the critical southwestern antennas.  Table~\ref{tab:obs} summarizes
the observing dates, frequencies, and participating antennas.

At all frequencies, a simple Gaussian model of Sgr A* with the size predicted by \citet{2006ApJ...648L.127B} was used as a starting model for the highest-quality epoch at that 
frequency.  After initial calibration, we concatenated data from all epochs for the Sgr A* field (separately for each frequency band) and iteratively imaged
and self-calibrated the combined datasets to generate high-quality clean component models.  These clean component models (1 per frequency)
were subsequently used for all observations at that frequency, with the position adjusted to the expected position of Sgr A* at that epoch based on the 
fit of \citet{2004ApJ...616..872R}.  These models were adjusted to the new and more accurate closure amplitude fits
at these wavelengths for Sgr A*
obtained below.  
Closure amplitudes were not used for the pulsar size because of the low SNR for the pulsar.  
In this way, the residual delay rates due to geometric model errors were minimised, and we obtain positions for Sgr A* and \sgr\ which
are approximately correct in the International Celestial Reference Frame (ICRF), with an absolutely positional uncertainty of $\sim 10$ milliarcseconds due
to the reference position and proper motion uncertainty of Sgr A* as given by \citet{2004ApJ...616..872R}.
The assumed proper motion of Sgr A* is $\mu_{Sgr A*}=(-3.151, -5.547)$ $\masy$ with position (17:45:40.0366,--29:00:28.217) at epoch MJD=56710.
  
After calibration, we performed for all epochs a visibility--by--visibility subtraction of the ungated dataset from the gated dataset as described in \citet{2014ApJ...780L...2B} to
remove the effect of Sgr A* from the image of \sgr.  A Gaussian model fit was performed in the visibility domain with the \verb+difmap+ package
\citep{1997ASPC..125...77S}, and errors on the size and position were estimated using the image-plane fitting task JMFIT in AIPS \citep{2003ASSL..285..109G}.  
In all cases, the best {\tt modelfit} result from \verb+difmap+ was consistent with the image-plane fit from JMFIT to within the errors.  We use the JMFIT results
and errors in the analysis below.
Table~\ref{tab:positions} shows the flux densities and observed positions for each epoch and band.  

Table~\ref{tab:sizes} reports measured apparent sizes for \sgr\ and Sgr A*.  Apparent sizes for \sgr\ were determined from 
image fits as described above with deconvolution of the synthesized beam.  
Sizes for Sgr A* were determined from closure amplitude fitting following the techniques described in \citep{2014ApJ...790....1B}.  
These provide the highest quality measurements of the size of Sgr A* at these wavelengths.

\begin{deluxetable}{rllllll}
\tabletypesize{\footnotesize}
\tablecaption{Observed Positions of \sgr\ \label{tab:positions}}
\tablehead{
\colhead{MJD} & \colhead{Band} & \colhead{Flux Density} & \colhead{ICRF RA} & \colhead{ICRF Dec.} & \colhead{$\Delta\alpha$} & \colhead{$\Delta\delta$}
\\
              &                &\colhead{(mJy)} &   \colhead{(J2000)} & \colhead{(J2000)} &  \colhead{(mas)} &  \colhead{(mas)} 
}
\startdata
    56422 &  X & 0.56 & 17 45 40.166377 $ \pm  0.000020 $ & -29 00 29.8960 $\pm 00.0002 $ & $ 1699.97 \pm    0.26 $ & $ -1683.38 \pm    0.20 $ \\ 
    56444 &  X & 0.76 & 17 45 40.166327 $ \pm  0.000012 $ & -29 00 29.8962 $\pm 00.0001 $ & $ 1699.50 \pm    0.16 $ & $ -1683.24 \pm    0.10 $ \\ 
    56473 & Ku & 0.58 & 17 45 40.166263 $ \pm  0.000008 $ & -29 00 29.8962 $\pm 00.0001 $ & $ 1698.91 \pm    0.10 $ & $ -1682.80 \pm    0.10 $ \\ 
    56486 &  X & 1.47 & 17 45 40.166222 $ \pm  0.000008 $ & -29 00 29.8965 $\pm 00.0001 $ & $ 1698.49 \pm    0.10 $ & $ -1682.90 \pm    0.10 $ \\ 
    56658 & Ku & 2.09 & 17 45 40.166264 $ \pm  0.000007 $ & -29 00 29.8960 $\pm 00.0001 $ & $ 1700.52 \pm    0.09 $ & $ -1679.79 \pm    0.10 $ \\ 
    \dots &  X & 1.18 & 17 45 40.166276 $ \pm  0.000010 $ & -29 00 29.8960 $\pm 00.0001 $ & $ 1700.68 \pm    0.13 $ & $ -1679.79 \pm    0.10 $ \\ 
    56710 & Ku & 1.07 & 17 45 40.166249 $ \pm  0.000005 $ & -29 00 29.8960 $\pm 00.0001 $ & $ 1700.77 \pm    0.07 $ & $ -1679.00 \pm    0.10 $ \\ 
    \dots &  X & 0.94 & 17 45 40.166237 $ \pm  0.000014 $ & -29 00 29.8962 $\pm 00.0002 $ & $ 1700.62 \pm    0.18 $ & $ -1679.20 \pm    0.20 $ \\ 
    56750 &  K & 0.92 & 17 45 40.166235 $ \pm  0.000004 $ & -29 00 29.8961 $\pm 00.0001 $ & $ 1700.94 \pm    0.05 $ & $ -1678.49 \pm    0.10 $ \\ 
    \dots &  Q & 0.54 & 17 45 40.166238 $ \pm  0.000003 $ & -29 00 29.8959 $\pm 00.0001 $ & $ 1700.98 \pm    0.04 $ & $ -1678.29 \pm    0.10 $ \\ 
    56772 &  X & 1.00 & 17 45 40.166246 $ \pm  0.000016 $ & -29 00 29.8958 $\pm 00.0002 $ & $ 1701.27 \pm    0.21 $ & $ -1677.86 \pm    0.20 $ \\ 
    \dots & Ku & 1.22 & 17 45 40.166204 $ \pm  0.000008 $ & -29 00 29.8961 $\pm 00.0001 $ & $ 1700.72 \pm    0.10 $ & $ -1678.16 \pm    0.10 $ \\ 
    56892 & Ku & 0.63 & 17 45 40.166196 $ \pm  0.000028 $ & -29 00 29.8959 $\pm 00.0003 $ & $ 1701.65 \pm    0.37 $ & $ -1676.15 \pm    0.30 $ \\ 
    56899 &  K & 0.26 & 17 45 40.166250 $ \pm  0.000025 $ & -29 00 29.8959 $\pm 00.0003 $ & $ 1702.42 \pm    0.33 $ & $ -1676.06 \pm    0.30 $ \\ 
    \dots &  Q & 0.15 & 17 45 40.166215 $ \pm  0.000005 $ & -29 00 29.8958 $\pm 00.0001 $ & $ 1701.95 \pm    0.07 $ & $ -1675.90 \pm    0.10 $ \\ 
\enddata
\end{deluxetable}

\begin{deluxetable}{llllllllll}
\tabletypesize{\footnotesize}
\tablecaption{Apparent Sizes of \sgr\ and Sgr A* \label{tab:sizes}}
\tablehead{
              &                & \multicolumn{3}{c}{\sgr} & \multicolumn{3}{c}{Sgr A*} \\ 
\colhead{MJD} & \colhead{Band} & \colhead{$b_{maj}$} & \colhead{$b_{min}$} & \colhead{$b_{pa}$} & \colhead{$b_{maj}$} & \colhead{$b_{min}$} & \colhead{$b_{pa}$} \\ 
              &                &  \colhead{(mas)}    &  \colhead{(mas)}    &  \colhead{(deg)}    &  \colhead{(mas)}    &  \colhead{(mas)}    &  \colhead{(deg)}    
}
\startdata
56422 & X & $ 16.9^{+ 0.9}_{-0.9} $ & $  9.1^{+ 1.2}_{-1.3} $ & $  81.4^{+  6.3}_{ -5.9} $ & $ 17.24^{+ 0.05}_{- 0.07} $ & $  9.00^{+ 0.32}_{- 0.40} $ & $  82.0^{+  0.9}_{-  0.7} $ \\ 
56444 & X & $ 15.0^{+ 0.6}_{-0.6} $ & $  5.2^{+ 1.2}_{-1.4} $ & $  88.8^{+  3.7}_{ -3.4} $ & $ 16.92^{+ 0.20}_{- 0.20} $ & $  8.40^{+ 0.56}_{- 0.52} $ & $  81.6^{+  1.9}_{-  1.2} $ \\ 
56473 & Ku & $  5.4^{+ 0.5}_{-0.5} $ & $  2.3^{+ 1.0}_{-2.0} $ & $  78.2^{+  9.8}_{ -9.2} $ & $  5.42^{+ 0.02}_{- 0.02} $ & $  2.87^{+ 0.04}_{- 0.14} $ & $  81.8^{+  0.4}_{-  0.4} $ \\ 
56486 & X & $ 15.2^{+ 0.3}_{-0.3} $ & $  8.5^{+ 0.5}_{-0.5} $ & $  78.2^{+  3.0}_{ -2.5} $ & $ 16.88^{+ 0.17}_{- 0.16} $ & $  8.04^{+ 1.00}_{- 0.68} $ & $  81.2^{+  2.9}_{-  1.9} $ \\ 
56658 & Ku & $  4.9^{+ 0.6}_{-0.5} $ & $  3.0^{+ 1.1}_{-1.7} $ & $  78.2^{+ 17.3}_{-15.7} $ & $  5.38^{+ 0.07}_{- 0.03} $ & $  2.52^{+ 0.07}_{- 0.08} $ & $  81.8^{+  1.0}_{-  1.2} $ \\ 
\dots & X & $ 16.3^{+ 0.7}_{-0.7} $ & $  8.5^{+ 1.1}_{-1.3} $ & $  82.1^{+  4.8}_{ -4.4} $ & $ 16.88^{+ 0.31}_{- 0.29} $ & $  8.64^{+ 0.64}_{- 0.60} $ & $  86.4^{+  1.2}_{-  2.4} $ \\ 
56710 & Ku & $  5.0^{+ 0.3}_{-0.3} $ & $  2.1^{+ 0.7}_{-1.1} $ & $  84.2^{+  5.9}_{ -6.7} $ & $  5.43^{+ 0.01}_{- 0.02} $ & $  2.70^{+ 0.09}_{- 0.10} $ & $  81.8^{+  0.4}_{-  0.2} $ \\ 
\dots & X & $ 17.2^{+ 0.6}_{-0.6} $ & $  8.0^{+ 0.9}_{-0.9} $ & $  84.3^{+  3.5}_{ -3.2} $ & $ 17.04^{+ 0.12}_{- 0.12} $ & $  8.04^{+ 0.44}_{- 0.48} $ & $  81.6^{+  0.7}_{-  0.4} $ \\ 
56750 & K & $  2.6^{+ 0.3}_{-0.2} $ & $  1.4^{+ 0.5}_{-0.9} $ & $  86.5^{+ 14.0}_{-12.2} $ & $  2.68^{+ 0.00}_{- 0.00} $ & $  1.45^{+ 0.03}_{- 0.03} $ & $  81.8^{+  0.0}_{-  0.0} $ \\ 
\dots & Q & $  0.5^{+ 0.2}_{-0.5} $ & $  0.3^{+ 0.6}_{-0.3} $ & $  98.3^{+ 40.4}_{-41.2} $ & $  0.730^{+ 0.003}_{- 0.001} $ & $  0.41^{+ 0.02}_{- 0.01} $ & $  80.0^{+  0.6}_{-  1.2} $ \\ 
56772 & X & $ 15.4^{+ 0.9}_{-0.5} $ & $  9.9^{+ 0.8}_{-1.5} $ & $  80.1^{+  5.2}_{-10.6} $ & $ 17.00^{+ 0.11}_{- 0.11} $ & $  8.28^{+ 0.68}_{- 0.64} $ & $  81.6^{+  0.8}_{-  0.5} $ \\ 
\dots & Ku & $  5.5^{+ 0.5}_{-0.5} $ & $  3.6^{+ 0.8}_{-1.0} $ & $  78.9^{+ 12.8}_{-13.1} $ & $  5.38^{+ 0.02}_{- 0.02} $ & $  2.71^{+ 0.10}_{- 0.13} $ & $  81.8^{+  0.2}_{-  0.4} $ \\ 
56892 & Ku & $  7.0^{+ 1.9}_{-2.1} $ & $  2.3^{+ 3.1}_{-2.3} $ & $  77.0^{+ 28.0}_{-24.0} $ & $  5.41^{+ 0.01}_{- 0.01} $ & $  2.71^{+ 0.07}_{- 0.07} $ & $  81.8^{+  0.2}_{-  0.2} $ \\ 
56899 & K & $  3.6^{+ 1.1}_{-3.6} $ & $  4.0^{+ 0.1}_{-4.0} $ & $  95.0^{+ 15.0}_{-37.0} $ & $  2.84^{+ 0.02}_{- 0.02} $ & $  2.16^{+ 0.01}_{- 0.12} $ & $  94.4^{+  2.4}_{-  2.4} $ \\ 
\dots & Q & $  <0.7$ & \dots & \dots & $  0.722^{+ 0.005}_{- 0.003} $ & $  0.14^{+ 0.06}_{- 0.04} $ & $  87.2^{+  0.6}_{-  1.2} $ \\ 
\hline 
 Avg. & X & $ 15.6^{+ 1.3}_{ 0.4} $ & $  8.4^{+ 0.8}_{ 1.7} $ & $  82.5^{+  4.5}_{-  3.5} $ & $ 17.16^{+ 0.05}_{- 0.05} $ & $  8.64^{+ 0.28}_{- 0.28} $ & $  82.0^{+  0.7}_{-  0.5} $ \\ 
 Avg. & Ku & $  5.2^{+ 0.4}_{ 0.2} $ & $  2.8^{+ 0.7}_{ 0.6} $ & $  81.4^{+  2.8}_{-  3.2} $ & $  5.41^{+ 0.01}_{- 0.01} $ & $  2.62^{+ 0.05}_{- 0.05} $ & $  81.8^{+  0.2}_{-  0.2} $ \\ 
 Avg. & K & $  2.6^{+ 1.0}_{ 0.0} $ & $  1.7^{+ 2.3}_{ 0.3} $ & $  88.2^{+  6.8}_{-  1.7} $ & $  2.70^{+ 0.01}_{- 0.01} $ & $  1.58^{+ 0.06}_{- 0.06} $ & $  83.6^{+  0.6}_{-  0.6} $ \\ 
 Avg. & Q & $  0.6^{+ 0.1}_{ 0.1} $ & $  0.0^{+ 0.3}_{ 0.0} $ & $  96.9^{+  1.4}_{-  6.9} $ & $  0.730^{+ 0.003}_{- 0.003} $ & $  0.36^{+ 0.03}_{- 0.01} $ & $  81.8^{+  0.6}_{-  1.2} $ \\ 
\enddata
\end{deluxetable}

\section{Astrometric Results for \sgr}
\label{sec:astrometry}

\begin{figure}[p]
\includegraphics{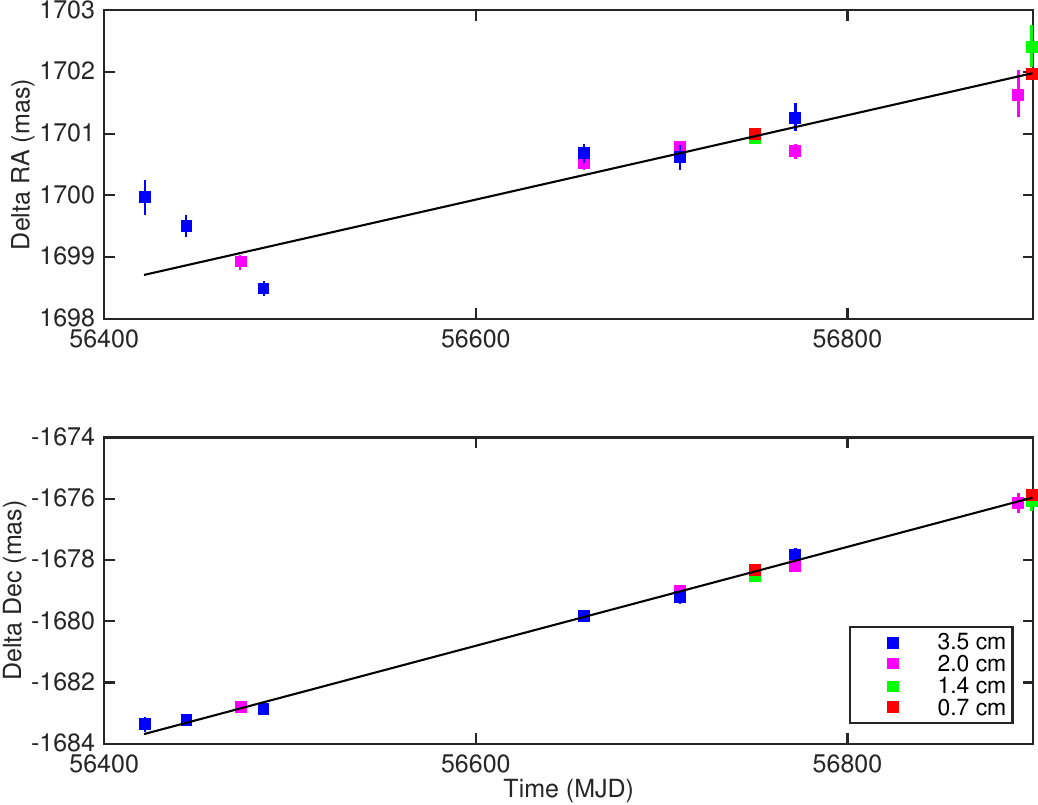}
\caption{Position as a function of time for \sgr\ relative to Sgr A*.  Different colors are used to identify the
wavelength of observations.  The solid black line shows the best-fit proper motion in each coordinate (see text for details
of the fit).
\label{fig:positions}
}
\end{figure}

We show positions as a function of time in Figure~\ref{fig:positions}.  
We fit the proper motion, acceleration, and core shift to the astrometric position of \sgr\ using the following equations:
\begin{eqnarray}
\Delta\alpha & = & \Delta\alpha_0 + \mu_\alpha * ({\rm MJD - MJD_0}) + {1 \over 2} a_\alpha * ({\rm MJD - MJD_0})^2 - \Phi_\alpha \lambda \\
\Delta\delta & = & \Delta\delta_0 + \mu_\delta * ({\rm MJD - MJD_0}) + {1 \over 2} a_\delta * ({\rm MJD - MJD_0})^2 - \Phi_\delta \lambda \nonumber, 
\label{eqn:astrofit}
\end{eqnarray}
where ${\rm MJD_0 = 56686}$ is the mid-point of our observations.  Solutions are calculated for
proper motion only ($a=0, \Phi=0$), proper motion and acceleration ($\Phi=0$), and proper motion and core shift ($a=0$).
Results are tabulated in Table~\ref{tab:fits}.  Note that the sign on the final term is defined to reflect that the core shift is
reflective of an actual shift in the position in Sgr A*.  
Fits were performed using a weighted least-squares method for each solution with
errors for each data point rescaled such that the reduced $\chi^2$ was equal to one.

\subsection{Astrometric Error Analysis}

\begin{figure}[p]
\includegraphics{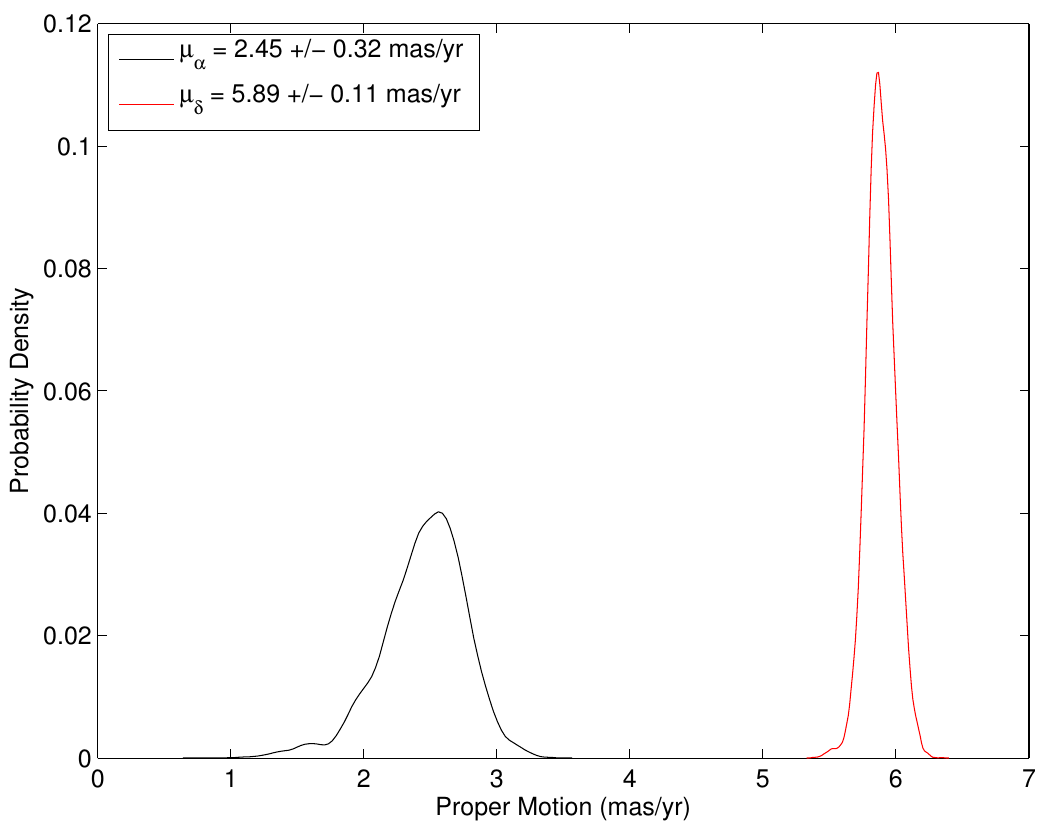}
\caption{Empirical probability density distribution of proper motion solutions
from the bootstrap resampling method.
\label{fig:bootstrap}
}
\end{figure}

As shown in Table~\ref{tab:fits}, inclusion of constant acceleration (due to Sgr A*) and/or constant core shift terms do not return detections of either of those parameters or significantly alter the quality of the fits.   Accordingly, we consider here the proper-motion-only fits in order to assess the presence of systematic errors in our measured \sgr\ positions.  Using all of the data and the formal positional errors estimated from JMFIT, we calculate $\chi^2$ values for each fit in right ascension and declination. None of the $\chi^2$ values are consistent with a good fit and accurately estimated errors. In particular, errors in declination are underestimated by $\sim$25\% while errors in right ascension are underestimated by a factor of $\sim$3.   As shown in Figure~\ref{fig:positions}, there is an apparent systematic error in the right ascension position associated with the first four epochs which accounts for most of the increase in $\chi^2$.  
A number of potential effects could cause this discrepancy, and we examine each of them in turn.

\begin{deluxetable}{llrrrr}
\tablecaption{Proper Motion Fits for \sgr\ \label{tab:fits}}
\tablehead{
\colhead{Parameter} & \colhead{Units} & \colhead{PM Bootstrap} & \colhead{PM LSQ} & \colhead{PM + Accel.} & \colhead{PM + Core Shift} 
}
\startdata
$\Delta\alpha_0$ & (mas)    & $ 1700.53 \pm   0.07 $ & $ 1700.52 \pm   0.07 $ &$ 1700.54 \pm   0.09 $ &$ 1700.52 \pm   0.19 $  \\ 
$\Delta\delta_0$ & (mas)    & $ -1679.41 \pm   0.03 $ & $ -1679.41 \pm   0.03 $ &$ -1679.42 \pm   0.05 $ &$ -1679.35 \pm   0.10 $  \\ 
$\mu_\alpha$ & (mas/yr)      & $ 2.45 \pm 0.33 $ & $ 2.50  \pm 0.22 $ & $ 2.49  \pm 0.23 $ & $ 2.51  \pm 0.31 $ \\ 
$\mu_\delta$ & (mas/yr)      & $ 5.88 \pm 0.11 $ & $ 5.90  \pm 0.09 $ & $ 5.92  \pm 0.10 $ & $ 5.84  \pm 0.12 $ \\ 
$a_\alpha$ & (mas/yr$^2$)      & \dots & \dots & $ -0.2  \pm 1.0 $ & \dots \\ 
$a_\delta$ & (mas/yr$^2$)      & \dots & \dots & $ 0.2  \pm 0.5 $ & \dots \\ 
$\Phi_\alpha$ & (mas/cm)      & \dots & \dots & \dots & $ -0.00  \pm 0.10 $ \\ 
$\Phi_\delta$ & (mas/cm)      & \dots & \dots & \dots & $ 0.03  \pm 0.05 $ \\ 
$ \chi^2_\alpha /d.o.f._\alpha$ &  & \dots & $ 112.3 / 13 $ &$ 111.8 / 12 $ &$ 112.3 / 12 $ \\ 
$ \chi^2_\delta /d.o.f._\delta$ &  & \dots & $ 17.0 / 13 $ &$ 16.8 / 12 $ &$ 16.5 / 12 $ \\ 
\enddata
\end{deluxetable}

The first three possibilities concern calibration.  As noted in \citet{2014ApJ...790....1B}, it is difficult to generate an accurate model of Sgr A* to use for calibration, because scatter-broadening causes it to be heavily resolved on most VLBA baselines.  This problem is doubly severe in right ascension compared to declination, as the shorter VLBA baselines are predominantly east-west (meaning higher resolution in the right ascension coordinate) and the major axis of Sgr A* scattering is almost aligned with the right ascension axis.  An incorrect model of Sgr A* will lead to different positional offsets between epochs if the $(u,v)$ coverage differs (which it does, due to different failed antennas), as the calibration changes the data in order to match the incorrect model.  In this sense the first 4 epochs are not especially poor, nor should this type of error lead to a right ascension error which apparently varies linearly with time.  It would affect the 8 GHz observations more severely than 15 GHz due to the more severe effects of scattering, 
and a higher proportion of the first four epochs are at 8 GHz, so this could simply be random errors giving the appearance of linear motion in the first four epochs.

Second, the calibration of the phased VLA could be worse in the first 4 epochs.  During this time, we used the source NRAO 530 to determine the real-time solutions to phase up the VLA, and after this time we used Sgr A* itself.  The angular offset from our target field to NRAO 530 is 16 deg, and so it is likely that the phased VLA gain is slightly lower and more time variable in these epochs.  Since the VLA heavily affects the calibration solution, being the most sensitive antenna, errors in its gain calibration could propagate through to larger errors in the other telescopes gain calibrations and hence positional offsets.  There is a direct link to the first 4 epochs, however, there is once again no reason to expect an apparently linear motion with time.

Third, ionospheric effects can introduce wavelength dependent astrometric errors.  In the case of our experiment, the small
separation, $\delta r$, between calibrator and target leads to a correspondingly small error in relative positions
\citep{2014ARA&A..52..339R}.  After the application
of total electron content (TEC) models as part of our calibration, residual ionospheric delay could correspond to $\lsim 10$ cm
at 3.5 cm.  This corresponds to an error of $\sim 3$ beamwidths, or $\delta\theta \lsim 10$ mas at 3.5 cm.  This leads to an
astrometric error for the 3.5 cm relative to a shorter wavelength (e.g., 7 mm) position where ionospheric effects are very small
of $\delta r \delta \theta \lsim  1\,\mu$as.   Thus, ionospheric effects do not affect our result.

Fourth, the apparent structure of Sgr A* itself could change with time.  Since we use a constant model at each frequency (derived from the concatenated datasets at that frequency) there is no way to detect or account for such a change, which could be due to intrinsic source effects (such as a time-variable core shift due to material propagating outwards along a jet) or time-variable scattering or scintillation.  By forcing Sgr A* back to the model position during calibration, we would impose an equivalent shift on the target magnetar.  Unlike the previous two explanations, evolution of Sgr A* could lead to a linear position change with time, and it would likely be along the right ascension axis, which is both the major axis for the scattering and for the intrinsic structure of Sgr A*.  However, the effect we see seems too large for this to be a likely explanation.  Various estimates of the scatter-broadened size of Sgr A* \citep{2006ApJ...648L.127B,2014ApJ...790....1B} show it to be constant over time within the error bars, and the intrinsic size of Sgr A* is likely much smaller than the required shifts here.  The specific case of a time-variable core shift is considered in more detail in Section~\ref{sec:sgra}.

Fifth, refractive wander could induce variations in the position of the size on an angular scale smaller than the size of the
scattered image.  The motion of $\sim 1$ mas in right ascension is much less than the image size of $\sim 15$ mas.  
The short 
timescale of the fluctuation $\sim 10^2$ days, however, appears inconsistent with the refractive timescale 
$\tau_R \sim D\theta/v \gsim 8\times 10^2$  days at 3.4 cm, where $D\sim 3$ kpc is the Earth to scattering screen distance, $\theta$ is the scatter-broadened
angular size, and $v \sim 100 \kms$ is the relative velocity of the screen perpendicular to the line of sight.  
$\tau_R$ is a minimum timescale for significant refractive changes because the turbulent medium may be uniform on scales larger
than $D\theta$ in the scattering screen. 
As noted above, the apparent size of Sgr A* has remained constant over the course of $\gsim 10$ yr, suggesting
that the refractive timescale is likely quite large.  
In addition, the similarity between the pulsar and Sgr A* images, discussed below, suggests that the timescale for refractive
changes could be as large as $R/v \approx 1000$ yr, where $R\sim 0.1$ pc is the separation between the two sources.
Further, previous attempts to detect positional wander in Sgr A* relative to background quasars at a separation of 0.5 deg
have not shown any effect on scales larger than 400 $\mu$as on time scales
ranging from $\sim 1$ hour to 1 month \citep{2008ApJ...682.1041R}.  Longer timescales were probed by observations of water masers 
in Sgr B2, which has a similar degree of scatter broadening to Sgr A* \citep{1988ApJ...330..817G}.  No positional wander was detected 
on a timescale of 6 months to a limit of 18 $\mu$as rms for maser spots spread over a region 4\arcsec\ in scale.  

On the other hand,the astrometric offsets we see are only a fraction of the total potential refractive image wander, 
and thus could potentially occur on (proportionately) shorter timescales than $\tau_R$.
In particular, the recent discovery of substructure in the image of Sgr A* at 1.3 cm  wavelength 
\citep{2014arXiv1409.0530G} suggests that refractive effects could impact the determiation of the centroid of
\sgr\ and the correct model for Sgr A*.  
Point-like emission with a flux density $\sim 1\%$ of the total flux density was found, effectively demonstrating
that Sgr A* resides in the ``average'' image regime \citep{1989MNRAS.238..963N,1989MNRAS.238..995G}.  
The characteristic time scale for
the average image regime ranges from the diffractive timescale $\sim 1$ sec to the time for an
interferometric resolution element to move by a single beam $\sim 10$ yr, i.e., comparable to
the refractive timescale.
The amplitude of refractive image wander relative to the scattered size in the case of a shallow
turbulent spectrum ($\alpha < 2$) will scale as the ratio of the substructure to peak flux densities.  That ratio is $\sim 100$
at 1.3 cm in the results from \citet{2014arXiv1409.0530G}, a factor of 10 smaller than the actual imaging errors at 3.4 and 2.0 cm.  
Steeper turbulent spectra can lead to more widely varying refractive image wander.
We also do not know whether refractive image wander would be independent between Sgr A* and \sgr\; if the image wander is common
between the two sources then there would be no astrometric effect in our data.
For a single epoch of our data at Ku band, we did fit a Gaussian scattering model plus point source and found no significant change in
the resultant position for \sgr.
More detailed study of the wavelength characteristics and 
time variability of this substructure can determine whether refractive image wander is a significant limitation
for astrometry of \sgr.

Finally, the change in proper motion in right ascension could be due to real acceleration of the magnetar, during a close encounter with a massive star (a wide companion in a highly elliptical orbit, or a chance encounter).  
To produce the apparent change in proper motion between the first four and the last four epochs would
require acceleration $\sim 1.5\,{\rm\,cm\,s^{-2}}$, equivalent to the effect of a massive star at a 
distance $\sim 10^{-3}$ pc, or 25 mas.  
No star is known to be this close to \sgr\, (see the discussion below).
We also consider this implausible, since the radial acceleration from such an encounter would 
also affect the pulsar timing, and no large deviations from the long-term timing trends are seen around 
MJD 56500.  In particular, for an edge-on orbit we require a change in the period derivative of $\sim 10^{-10}$, an
order of magnitude larger than the period derivative $\dot{P}=6.12 \times 10^{-12}$ measured by \citet{2014ApJ...786...84K} at this epoch. 
In fact, timing observations detected an abrupt change in the period derivative at MJD=56450,
in the middle of the apparently linear proper motion and, therefore, in conflict with orbital motion.
If we take the conservative estimate of the measured period derivative as entirely due to acceleration, we set
an upper limit that the inclination angle must be $\lsim 4\deg$.
A more detailed analysis of the timing data will set even stronger constraints.
Thus, the encounter would have to had been seen almost face on, an unlikely but 
not impossible scenario.

Ultimately, we cannot confirm or rule out any of the explanations listed above with the exception of differential
ionospheric errors.  Additional astrometric epochs may eventually provide enough information to favour one of these explanations, but the additional uncertainty imposed on the right ascension proper motion due to the scatter in the first four epochs does not alter any of the conclusions which follow in this paper.  Removing this systematic uncertainty and approaching the S/N limited uncertainty in proper motion, however, will be crucial in detecting the acceleration of \sgr\ due to Sgr A* over a long period (see Section~\ref{sec:conclusions}).

Due to the relatively poor quality of the right ascension least-squares fit,
for the proper motion only case, we also performed a bootstrap resampling method
to estimate solutions and errors more accurately.  
The bootstrap method resampled
the data with replacement $10^4$ times.  We did not group the data by date, which may
permit some small amount of correlated error to propagate into our solutions, i.e., as the
result of same-day tropospheric corrections.
In practice, we found negligible difference in the solutions (i.e., $\delta \mu_\alpha =0.015 \masy$)
if we averaged the data by date before including it in the bootstrap algorithm.
We adopt the bootstrap result, which is consistent with the least-squares fit but with larger
errors, as our best result.

\subsection{The Proper Motion and Acceleration of \sgr}

We consider initially the astrometric fits to proper motion alone. The acceleration constraints are weak 
and the proper-motion only fits provide the highest accuracy.  
The best-fit proper motion is $\mu_\alpha=2.44 \pm 0.33$ $\masy$ and $\mu_\delta=5.89 \pm 0.11$ $\masy$ (67\% confidence limits).  
The  empirical probability distribution of $\mu_\delta$ is strongly
peaked and resembles a Gaussian (Figure~\ref{fig:bootstrap}).  The probability
distribution of $\mu_\alpha$, on the other hand, is asymmetric with
a longer tail to smaller values.  The 95\% confidence intervals for
the best-fit parameters are $1.6 < \mu_\alpha < 3.0$ $\masy$ and 
$5.7 < \mu_\delta < 6.1$ $\masy$.  

The best-fit proper motion is 
the opposite sign and within 25\% of the magnitude of the proper motion of Sgr A* relative to the ICRF (caused by Galactic
rotation).  This produces the result
that the pulsar is nearly at rest relative to the ICRF.  Nevertheless, it is the motion of the pulsar relative to 
Sgr A*, which is relevant for understanding its origin and future.
The transverse velocity for the pulsar relative to Sgr A* is $v=236 \pm 11\, \kms$ in position angle $22 \pm 2$ deg East-of-North.
From our fit, the positional offset of \sgr\ from Sgr A* at MJD 56406 (when it was first detected) was 2392 mas, corresponding to a distance of 0.097 pc. 
The observed separation of the pulsar is consistent with lower precision X-ray estimates \citep{2013ApJ...775L..34R}.

\begin{figure}
\includegraphics{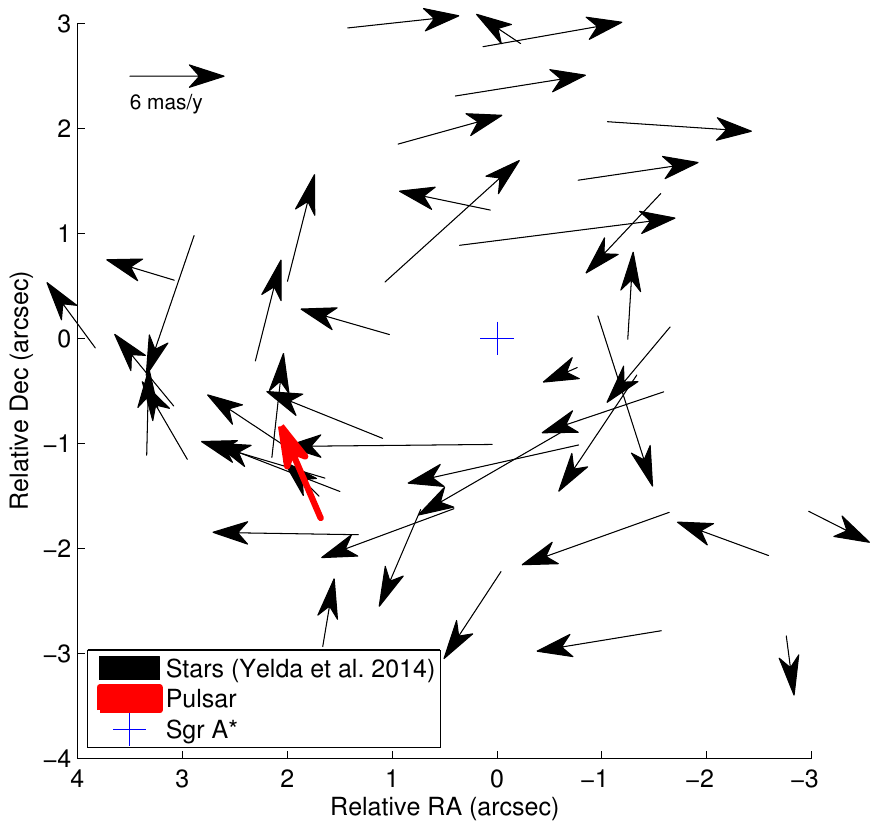}
\caption{Proper motion of the pulsar along with stars from \citet{2014ApJ...783..131Y}.
\label{fig:pmstars}
}
\end{figure}

The proper motion of the pulsar is similar to the motion of stars near Sgr A* studied through 
NIR adaptive optics \citep[Figure~\ref{fig:pmstars};][]{2014ApJ...783..131Y}.  The stars S2-4 and S2-6 are within 500 mas of \sgr\ and have 
proper motions of $(7.9,3.1)$ $\masy$ and $(7.9,2.3)$ $\masy$, respectively.  
The magnitude of the velocity vector is similar to 
that of these nearby Galactic Center stars,  however, the vector orientation differs by $\sim 45$ deg from 
that of the nearest stars.  This variation in the orientation is comparable to the variations among the stars.  
Thus, the motion of the pulsar appears to 
be consistent with clockwise (CW) rotation around Sgr A*.  In the next section, we explore the probability of
an origin in the CW stellar disk through Monte Carlo simulations and conclude 
that an origin in that disk is likely, although we cannot rule out an origin from the isotropic distribution of stars. 

Our fits are consistent with no acceleration with $3\sigma$ upper limits of 3 mas\,yr$^{-2}$ and 1.5 mas\,yr$^{-2}$ in each coordinate, respectively.
This does not provide a strong constraint because the maximal acceleration that can be obtained at this separation from Sgr A* is
0.04 mas\,yr$^{-2}$.  The magnitude of the acceleration is dependent on the distance $z$ along the line of the sight of the pulsar 
from Sgr A*.  If the pulsar remains bright at high frequencies for $\sim$ 3 -- 10 yr, future astrometric 
observations will have the sensitivity to detect the acceleration and, therefore, determine $z$.
Constraints on acceleration will grow rapidly with increasing observing time.

\begin{figure}[p]
\includegraphics{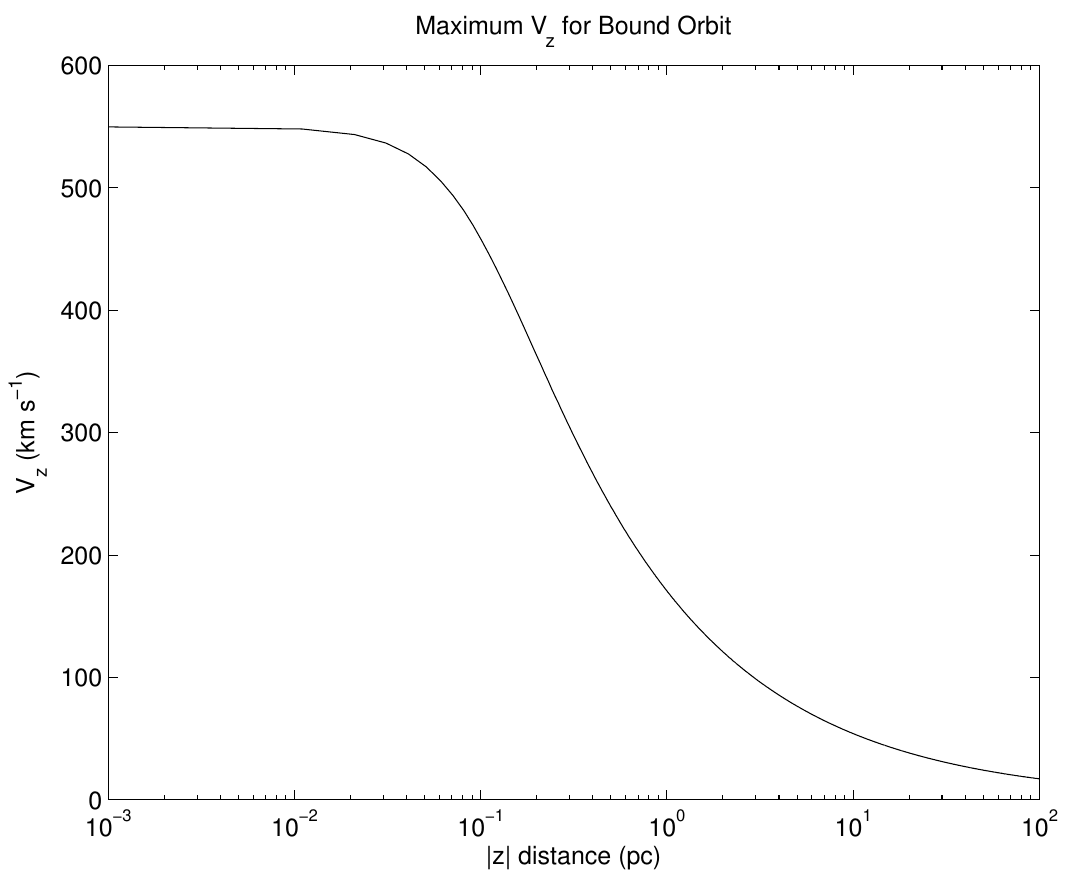}
\caption{Maximum $v_z$ as a function of $|z|$ in order for the pulsar to be bound to Sgr A*.
\label{fig:boundorbit}
}
\end{figure}

In the absence of an acceleration, we cannot determine conclusively whether the pulsar is bound to Sgr A*.  We can
set limits on the velocity along the line of sight, $v_z$, as a function of the line of sight distance from the pulsar
to Sgr A*, $z$, for which the pulsar is bound using
the constraint that $E=T+U < 0$ (Figure~\ref{fig:boundorbit}).  For $|z| < 0.1 (1.0)$  pc, velocities up to 550 (170) $\kms$ are bound
to Sgr A*. \citet{2006ApJ...643..332F} estimate the mean natal three-dimensional velocity for pulsars at 380 $\kms$, corresponding
to a single coordinate mean of 220 $\kms$.  Therefore, to $|z| \lsim 1.0$ pc, the pulsar is likely to be bound.
For $z=0$ and $v_z=0$, the orbital period is $\sim 700$ y, perigee is $\sim 0.01$ pc and apogee is $\sim 0.1$ pc.
For higher velocities and values of $z$, the orbital period increases.  

Two other pieces of evidence support a hypothesis that the pulsar is bound to Sgr A*.  One, the chance of random superposition
of an unbound object within 3\arcsec of Sgr A* in the 1400\arcsec-square Swift field of view is $\sim 10^{-5}$.  Two, the relatively
high transverse proper motion gives a short window $\sim 400$ yr in which the pulsar will be within 0.1 pc of Sgr A*.  This
timescale is shorter than the likely lifetime of the pulsar $\sim 10^3$ -- $10^4$ yr.  Thus, the observed position and velocity 
point to an object that is likely bound to Sgr A*.

We also set an upper limit to the parallax with a fit to the ICRF position.  We find 
$\pi < 0.6$ mas at 95\% confidence.  The limit is poor due to the relatively limited 
temporal coverage and due to the systematic errors in the right ascension position.
Future observations will significantly improve this limit.

\subsection{Monte Carlo Simulations of the Pulsar Origin}


\begin{figure}
  \centering \includegraphics[width=\textwidth]{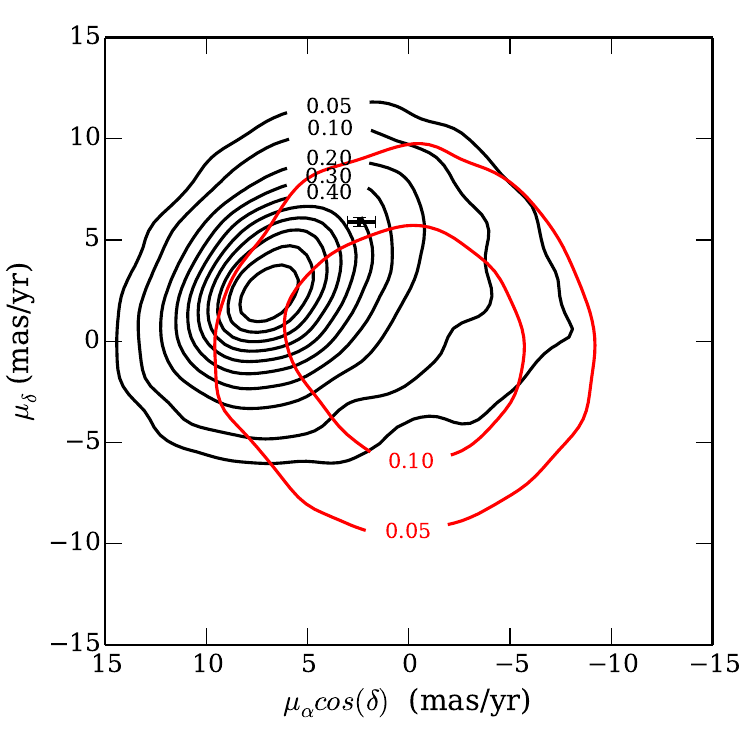}
\caption{ 
 The proper motion distribution of pulsars kicked from the disk
({\em black lines}) and kicked from an isotropic distribution ({\em
red lines}). Each line shows a contour of the two-dimensional
probability density of proper motions for stars that are near the
position of PSR J1745-2900.  Nearly all the stars that originate in the
disk remain on clockwise orbits around Sgr A*. The pulsar's natal kick
causes the distribution to be much broader than the observed disk of
massive stars. The probability density distribution of stars that
originate in the isotropic distribution is broad, and so an isotropic
origin can not be excluded.
The observed proper motion of  \sgr\ and our 95\%
confidence intervals are shown with the black symbol and error bars.
\label{fig:pm} }
\end{figure}

\begin{figure}
  \centering \includegraphics[width=\textwidth]{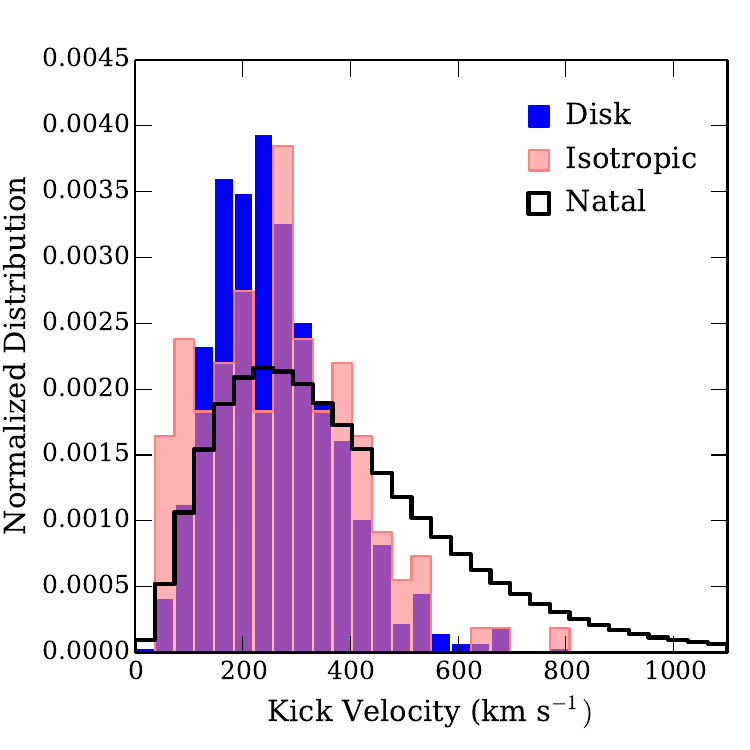}
\caption{
Kick velocity distribution. The natal kick velocity
distribution assumed in this work is shown in black. The solid blue
histogram shows the kick velocity distribution of pulsars that
originated in the disk and have a similar position and proper motion
to \sgr; the same distribution for stars originating in the
isotropic distribution is shown in red. For the pulsar to have originated in the disk or isotropic cloud
it would
require a kick of $100 - 500\,{\rm km}\,{\rm s}^{-1}$.
\label{fig:kick} }
\end{figure}

\begin{figure}
  \centering \includegraphics[width=\textwidth]{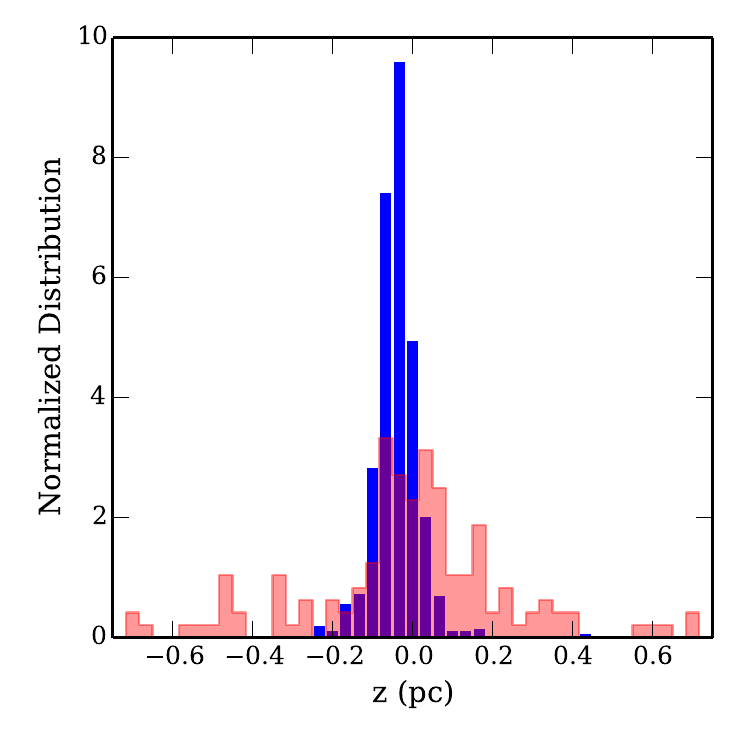}
\caption{ 
Distribution of line-of-sight distance ($z$) relative to Sgr A*. The
solid blue histogram shows the distance for stars that were
kicked out of the CW stellar disk. The red transparent histogram shows
the distribution for stars that originated from an isotropic
distribution. In both cases, we only show the position of pulsars with
similar kinematic properties to \sgr. Pulsars that
originated in the disk are centrally concentrated with $|z| \lesssim
0.1\,$pc in 88 percent of the simulations.
\label{fig:z} }
\end{figure}

The observed proper motion of \sgr\ is CW, consistent with
the young disk of massive stars orbiting Sgr A*
\citep{2006ApJ...643.1011P,2009ApJ...690.1463L,2014ApJ...783..131Y}, however, the
angle of its orbit is offset by $\approx 45\deg$.  We assess if a
natal kick can move a pulsar onto this orbit using a suite of Monte
Carlo simulations of potential orbits of pulsars in the Galactic
Center assuming it originated in the stellar disk.  
We compare these
results to Monte Carlo simulations of an alternative model of the pulsar origin, 
where the pulsars are kicked from an
isotropic distribution of stars with a thermal distribution of
eccentricities.

For the stars originating in the disk, we initially generate orbits
consistent with the best fit parameters of the observed CW disk
\citep[$\Omega \approx 96\pm 3\deg$, $i \approx 130 \pm 8\deg$, $e
  \approx 0.27 \pm.07$;][]{2014ApJ...783..131Y}. We then kick each star
from its orbit by selecting a random velocity from the
double-exponential distribution of \citet{2006ApJ...643..332F},
which has a median kick velocity of $\approx 380\kms$. 
Other distributions of pulsar kick velocities have been created with
similar broad properties \citep[e.g.,][]{2002ApJ...568..289A}.
Although this
distribution was derived to fit the overall pulsar (not magnetar) population of the Milky Way, it is
consistent with the observed distances of magnetars that lie near massive star forming
regions and the two measured magnetar transverse velocities \citep{2007ApJ...662.1198H,2012ApJ...748L...1D}.
We can compare these results with stars selected from an isotropic distribution of orbits with  a thermal eccentricity distribution and
three-dimensional density profile $n \propto r^{-2}$ consistent with
the stars not located in the disk
\citep{2006ApJ...643.1011P,2013ApJ...764..155L,2014ApJ...783..131Y}.  
In both
models, $\sim 10^7$ orbits are each integrated for a uniform random duration up to $10^4\,$yr. The longest integration
time is comparable
to the inferred spin-down age of \sgr, and approximately ten orbital periods at $0.1\,$pc from Sgr A*; it is also
slightly less than the Newtonian precession timescale at that distance
\citep{2011MNRAS.412..187K}.  As the Newtonian precession only acts in
the plane of the orbit (i.e. keeps the inclination fixed), it has
little impact on the currently observable properties of the magnetar,
so longer integration times should not be necessary.  Shorter
integration times are needed to assess unbound orbits which spend
little time within 0.1 pc of Sgr A*.  
We found no noticeable difference with the resultant distributions in 
using a fixed integration time of $10^4$ or $10^5\,$yr.
These integrations are carried out
using the {\sc galpy}\footnote{https://github.com/jobovy/galpy} code with a static potential that replicates the
Milky Way rotation curve from the Galactic Center through the halo.

In Figure~\ref{fig:pm}, we compare the observed proper motion of
\sgr\ with the Monte Carlo simulations by selecting only pulsars within $1\,\arcsec$ of the position of \sgr.  Stars that are kicked out of the stellar disk tend to remain on CW orbits. The observed proper motion of \sgr\ is completely consistent with this distribution.   The proper
motion distribution of the isotropic origin, on the other hand,  is rather broad, which
prevents us from excluding it as the origin at a robust level.
 Assuming the disk fraction of stars is $50\,\%$, we find that $83\,\%$ of the pulsars with kinematic properties similar to \sgr\ were born in the CW disk.  However, \citet{2014ApJ...783..131Y} has recently revised the estimated disk fraction down to 
$25\,\%$. Using this lower disk fraction as our prior, lowers the likelihood of originating form the disk to $\approx 62\,\%$.  

Our simulations support the hypothesis that stars originating in the disk and isotropic distribution 
are likely to remain bound to Sgr A*.  
For stars originating in the disk, with the ordinary pulsar kick
distribution, none of $\sim$720 stars with kinematic properties similar to
the magnetar were unbound to Sgr A* and central pc of stars.
For stars originating from the isotropic distrbution, there was 1 star
out of 149 with similar kinematic properties that was unbound to the central parsec.  It received a kick
of 675 \kms, and was 172 years old.

Unlike in isolated massive star clusters, 
only very large natal kicks are sufficient
to unbind a pulsar from Sgr A*.  
In Figure~\ref{fig:kick}, we show the
natal kick velocity distribution assumed in this work, to the
distribution of kick velocities that resulted in magnetars with a
position and proper motion similar to \sgr.  We find that the range
of kicks allowed, without unbinding the magnetar spans a large range
from $100-500\kms$. 
These limits are consistent with the finding for
the four other magnetars with observed kick velocities 
($\sim$ 130 -- 350 \kms) 
are comparable to that of ordinary pulsars \citep{2007ApJ...662.1198H,2012ApJ...748L...1D,2012ApJ...761...76T}.
Early models predicted that magnetar birth kicks could be much larger than those of normal pulsars, enabled by exotic kick mechanisms that could not operate efficiently in the formation of normal pulsars 
\citep{1992ApJ...392L...9D,1993ApJ...408..194T};
our results for PSR J1745-2900 add further evidence disfavouring these predictions.
On its own this one transverse velocity does not strongly constrain the natal kick velocity distribution of magnetars, but 
three measurements which each require unlikely viewing geometries to support a natal kick in excess of 500 \kms  strongly disfavour a distribution which peaks at high velocities ( $>500$ \kms).
On its own this one kick measurement does not provide much constraint on the natal kick velocity, but three kick
measurements in which the kick is either $<500$ \kms or the alignment is
very fortuitous is unlikely.
It is difficult to give a quantitative estimate, since the width and functional form of the velocity distribution 
would also be free parameters, but a simple assumption like the scaling up of the Maxwellian distribution assumed for normal pulsars 
to give a $2\times$ higher kick (and hence a mean 3d kick velocity of 800 \kms vs 400 \kms) gives a probability of
 sampling 5 magnetars with transverse velocity measurements $< 300$ \kms of $\lsim 0.1$\%.

In Figure~\ref{fig:z}, we show the expected distance of
\sgr\ relative to Sgr A* assuming it formed in a disk or from
an isotropic distribution.  At \sgr's current position, stars
in the CW stellar disk are nearest us and have measured
positive radial velocities.  We find that if the magnetar originated
in the disk, it has $|z| \lesssim 0.1\,$pc in $88\,\%$ of the
simulations.  If the magnetar originated from an isotropic
distribution of stars we find it still has $|z| \lesssim 0.1\,$pc and
a proper motion comparable to the measured value in $\approx 49\,\%$ of our simulations.
As noted above,
future interferometric observations of \sgr\ will likely detect the
acceleration of \sgr\ and determine $|z|$.

\section{Implications for Sgr A*}
\label{sec:sgra}

\subsection{Size Fits for \sgr\ and Sgr A*}

\begin{figure}[p]
\includegraphics{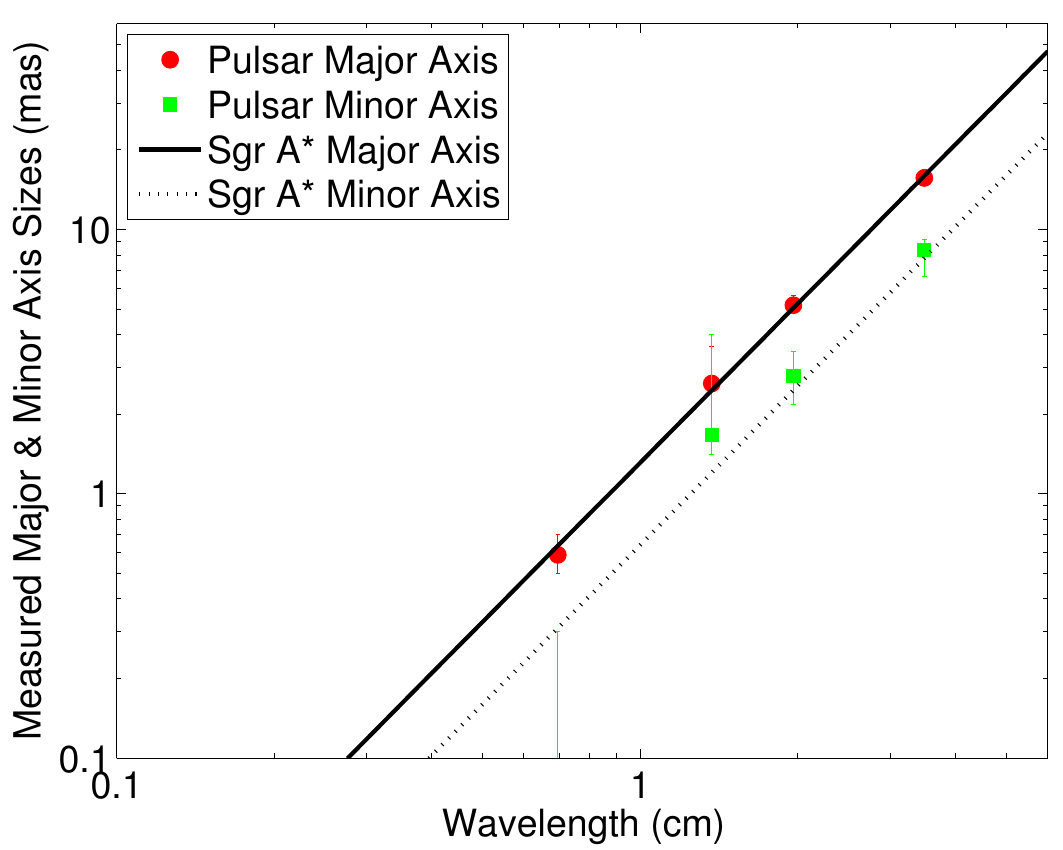}
\caption{Pulsar size as a function of wavelength for major and minor axes.  Data are plotted against
the fitted major and minor axis size of Sgr A*, which scales as $\lambda^2$.
\label{fig:size}
}
\end{figure}

We plot the observed size of the pulsar as a function of wavelength in Figure~\ref{fig:size}.  The pulsar size shows
excellent agreement from 3.6 cm to 0.7 cm with the scattering model for Sgr A* in both major and minor axes.  This result
further supports the angular broadening interpretation of \citet{2014ApJ...780L...2B} that the pulsar and Sgr A* share the same scattering
medium along the line of sight.  The similarity over all wavelengths is important because it demonstrates that the mean
scattering properties of either source will not change for a time scale $\tau \gsim R/v \sim 1000$ yr, where $R\sim 0.1$ pc is the separation
between the pulsar and Sgr A* and $v \sim 100 \kms$ is the velocity of the scattering screen across the line of sight.  
This timescale is much larger than the refractive timescale $\tau_R \sim 0.05$ yr at 7mm, indicating that the scattering
medium appears to be smooth over large scales.  

We also find new constraints on the size of Sgr A* from closure amplitude analysis of visibilities.  Closure
amplitude analysis provides the most accurate method for determining the size of Sgr A* \citep{2014ApJ...790....1B}.  We provide
new results for the size at 3.5 and 2.0 cm that are a substantial improvement on sizes previously measured  \citep{2004Sci...304..704}.
Improvements are due to the greater bandwidth and, therefore, improved sensitivity of VLBA observations and the inclusion
of the VLA which provides shorter baselines.  
At both wavelengths, the new measured major axis sizes are more precise and differ by less than
$2\sigma$ with the earlier measured sizes.  Any differences could be due to changes in 
the intrinsic size or in the refractive properties of the medium.
Including the new
3.5 cm major axis size with the earlier 6 cm VLBA size and L-band VLA sizes from \citet{2006ApJ...648L.127B}, we estimate
the major axis scattering normalization $b_{maj,scatt}=1.32 \pm 0.02$ mas\,cm$^{-1}$.  This is consistent within the errors with the
previous estimate of $b_{maj,scatt}$. The new minor axis sizes at 3.5 and 2.0 cm are a significant improvement over
previous values.  We combine these measurements with the previous 6 cm VLBA size but not the L-band VLA minor axis sizes, which were
corrupted by the presence of a radio transient, and find a new estimate of the minor axis scattering normalization, $b_{min,scatt}=0.67 \pm 0.02$ mas\,cm$^{-1}$.
The error is a factor of two smaller than previous estimates.  We also estimate the mean position angle from the 2 to 6 cm 
data to be $81.8 \pm 0.2$ deg.  These new estimates are more precise but within the errors of previous estimates 
and so do not affect significantly conclusions on the intrinsic size of Sgr A* \citep[e.g.,][]{2009A&A...496...77F,2014ApJ...790....1B}.

\subsection{Core Shift in Sgr A*}

The core shift arises in jet sources because the optical depth changes as a function of wavelength \citep{1979ApJ...232...34B,1995A&A...293..665F}.  At shorter
wavelengths, jets appear more compact and the $\tau=1$ surface is closer to the origin of the jet and, therefore,
the black hole.  We phase-reference the pulsed emission from \sgr\ to Sgr A* in this experiment.  The pulsed emission
must be intrinsically point-like since it originates within the pulsar magnetosphere, which has an
angular size smaller than 1$\mu$as.  Thus, any wavelength-dependence
in the position of the pulsar must originate from wavelength-dependent structures in Sgr A*.  
The discovery of refractive substructure by \citet{2014arXiv1409.0530G}, however, does suggest that differential image wander could also affect
this relative measurement.
Many accretion models for Sgr A*
suggest that the accretion flow is 
likely to be spherically symmetric and, therefore, produce zero core shift. Our data provides
a powerful opportunity to probe the structure of Sgr A* at wavelengths where the source image is strongly dominated
by interstellar scattering.  We define the core shift $\Phi$ in Equation~\ref{eqn:astrofit} for right
ascension and declination as a wavelength scaling to the position.

\begin{figure}[p]
\includegraphics{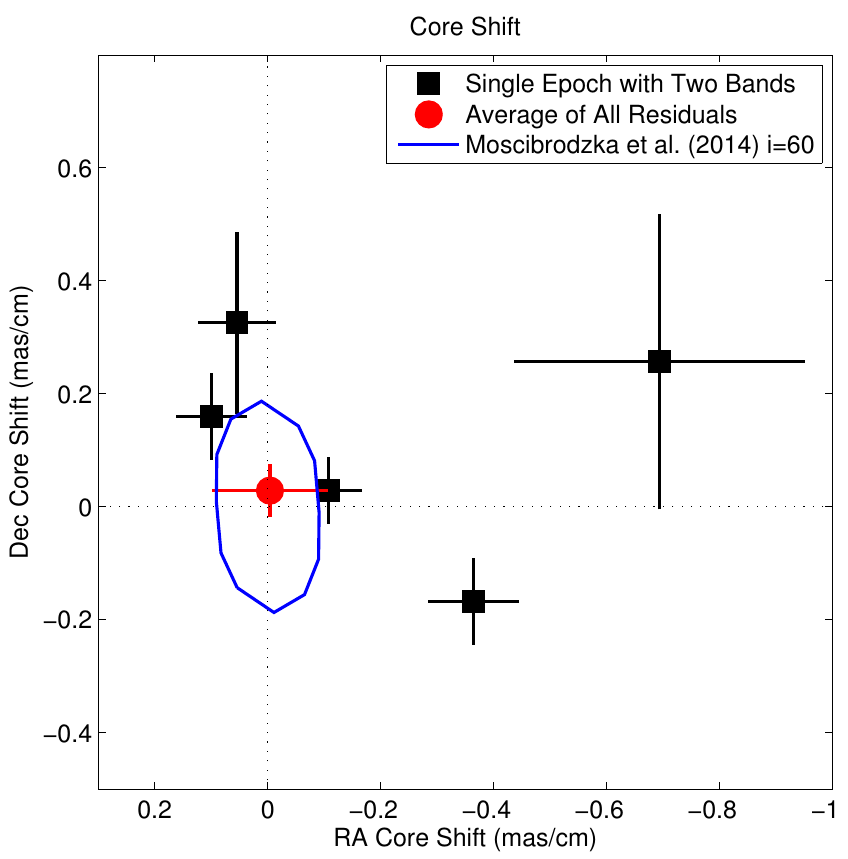}
\caption{Core shift for the four individual epochs with two frequency observations (black squares) and for the
average over all epochs (red circle).  The blue curve shows the core shift estimated from GRMHD jet simulations
with an inclination angle of 60 deg oriented towards different position angles.  Models are convolved
with the scattering beam and then the centroids are determined.  The measured and theoretical core
shifts agree in order of magnitude.
\label{fig:coreshift}
}
\end{figure}

As Table~\ref{tab:fits}
demonstrates, our least-squares fit does not find a significant detection of a constant core shift.  The $3\sigma$ upper limit is
$\sim 0.3$ mas\,cm$^{-1}$ in right  ascension and $\sim 0.2$ mas\,cm$^{-1}$ in declination.  The fit to all data is heavily weighted by the high SNR detections at wavelengths of 1.3 and 0.7 cm.
Giving equal weight to all epochs in the fit also does not lead to a statistically significant detection.
We also consider the possibility that the core shift may be time variable.  Sgr A* is known to have
time variable flux density at all wavelengths, including the longer wavelengths examined here \citep{2004AJ....127.3399H,2006ApJ...641..302M}.  
In five epochs (MJD 56658, 56710, 56750, 56772, and 56899), we obtain simultaneous observations at two wavelengths from which we can estimate a
core shift (Figure~\ref{fig:coreshift}).  
For three of these epochs, the two bands are Ku and X; in the fourth and fifth
epoch, the bands are K and Q.  In only one epochs (MJD 56772, Ku and X band) do
we see a $>3\sigma$ core shift with an
amplitude $\sim -0.4$ mas\,cm$^{-1}$; for epoch MJD 56899, K and Q band, the significance
is $\sim 2.8\sigma$ with a value of $0.7$ mas\,cm$^{-1}$.  
Errors in $\Phi$ are determined using the formal statistical errors on the measured positions rather than errors scaled to
achieve $\chi^2_\nu=1$, because we are testing the question of whether apparent systematics offsets may be due
to the core shift.
Given the large scatter in the core shift measurements
and systematic error in the right ascension positions, we conclude that we have produced an upper bound 
on the amplitude of the core shift.
Higher sensitivity measurements with three or more simultaneous frequencies are necessary to 
convincingly demonstrate the presence (or absence) of a core shift.
Multi-frequency measurements could also test against refractive effects due to large-scale inhomogeneities
in the scattering screen leading the positions for the two sources to wander independently.

The sign and angle of the coreshift is indicative of the direction of a jet.  
If our single epoch detection of the core shift is accurate,
this indicates a jet that extends primarily to the Southwest in position angle 245 deg East of North.  
\citet{2014ApJ...790....1B} found an extension of the source in position angle $\sim 95$ deg, which is not statistically 
consistent with the direction of the core shift.  
There is also some suggestion in the data of an alignment of the individual epochs along the 245 deg axis,
which could be consistent with a time-variable bipolar outflow.  

We compare the measurements with the core shift predictions presented in
\citet{moscibrodzka2014}.  The theoretical models are based on the general
relativistic magnetohydrodynamics simulations of an accreting black hole in
which jets are naturally produced by magnetic processes. These simulations
combined with the ray-tracing radiative transfer models can predict the
synthetic images of the jet-disk-black hole triad at various wavelengths
\citep{2013A&A...559L...3M,moscibrodzka2014}.
To compute the theoretical core
shifts, the time-averaged images of the jet models are convolved with the
scattering screen Gaussian for various position angles (PA) of the jet on the
sky. Then, the theoretical core shift is
$\Phi_{theory}(PA)=(\phi_1(PA)-\phi_2(PA))/(\lambda_1-\lambda_2)$, where
$\phi_{1,2}$ are the centroids of the scattering-broadened images at
$\lambda_{1,2}$=0.7 and 1.3 cm, respectively. The theoretical core shifts for
a jet inclined at $i=60$ deg and various PA are shown in Figure~\ref{fig:coreshift}.
We have chosen to show this model because its broadband synthetic spectrum and image size at mm-wavelength 
are consistent with all observations of Sgr A*.  Other self-consistent models 
may show different profiles.
We find that the model time-averaged core
shifts are of the same order of magnitude ($\sim 0.2\,{\rm mas\,cm^{-1}}$) as those measured only during
individual epochs of our observations. One can further test the time-dependent
GRMHD models of jets by studying the core shift as a function of time.  A
variable core shift puts strong constraints on the jet particle heating, 
its speed, and inclination angle but this requires more detailed
investigation.

We also consider a simple order of magnitude estimate of the amplitude of the core shift based on measured time lags
in variability for Sgr A*.  \citet{2006ApJ...650..189Y} observed delays of $\tau \sim 30$ minutes between variability at 1.3 and 0.7 cm.
If these delays represent the size of the photosphere at different wavelengths and the underlying
structure is a jet with velocity $v_j$ and inclination angle of 90 deg, 
then the apparent core shift is $\Phi \sim v_j\tau/D_{GC}/\Delta\lambda$
\citep{2009A&A...496...77F},   where $D_{GC}=8.3$  kpc  is the distance to the Galactic Center.
We find $\Phi\approx 0.7 (v_j/c)$ mas\,cm$^{-1}$, which is a comparable order of magnitude to the 
detailed model estimates and our measured values.

\section{Discussion and Conclusions}
\label{sec:conclusions}

We have shown that the magnetar most likely originated within the CW stellar disk orbiting Sgr A*, 
or, with less likelihood, the isotopic stellar population at a radius $\lsim 0.4$ pc.  
This statement rests on the assumption that the natal kick velocity distribution for magnetars
is similar to that of ordinary pulsars, which this result appears to support.
Both populations are a part of the Galactic Center cluster, which
consists of numerous O and WR stars with a characteristic age of $\sim 4$ Myr and a total mass $>10^4 M_\odot$; 
the most
massive existing stars have  initial masses up to $\sim 60 M_\odot$ \citep{2006ApJ...643.1011P,2013ApJ...764..155L}.
This supports the identification of magnetars as the remnants of supernova explosions of high mass stars.

Single stars appear to primarily create black holes for initial masses greater than $25 M_\odot$, but binarity
appears to create a pathway for greater mass loss leading to neutron star and possibly magnetar formation \citep{2002ApJ...578..335F,2008ApJ...685..400B}.
This was recently demonstrated by
the discovery of a potential binary companion to the progenitor of the magnetar J1647-45 in Westerlund 1 \citep[Wd 1;][]{2006ApJ...636L..41M,2013ApJ...763...82A,2014A&A...565A..90C}.  
Massive binary stars in the central few arcseconds of the Galactic Center cluster have been detected in three systems:  IRS 16SW,
IRS 16NE, and E-60  \citep[also known as S4-258;][]{2014ApJ...782..101P}.   E-60 and IRS 16SW are contact binaries, 
which may be necessary for the co-evolution of massive stars leading to a magnetar  \citep{2002ApJ...578..335F,2014A&A...565A..90C}.  
Co-evolution can lead to mass loss in the supernova progenitor, which reduces angular momentum loss
in the core during the final stages of stellar evolution and, thereby, enhancing the seed magnetic field required
for magnetar production.
The binary fraction among massive
stars in the Galactic Center cluster is estimated to be substantial and comparable to that seen in other dense clusters.
None of the known binary systems, however, are given a high probability of membership in the CW stellar disk
\citep{2014ApJ...783..131Y}; E-60 appears to be marginally bound or unbound to Sgr A*. 
Nevertheless, it is not unreasonable to argue that the pulsar originated in a massive 
binary system in the CW stellar disk.  
We note that one of the pulsars in the central 40 pc of the Galaxy, PSR J1746-2850II, also appears
to be a young, highly magnetized neutron star with a possible origin in the Arches or Quintuplet
cluster
\citep{2009ApJ...702L.177D}.
The other known pulsars are ordinary, long-period sources, whose origin could be in one of the young
stellar clusters or in other regions in the Galactic Center.

The association of the pulsar with an origin in the CW stellar disk supports the basic theory behind predictions
of the population of neutron stars, black holes, and pulsars in the Galactic Center.  The young stars in the Galactic Center
are expected to produce many compact objects as they evolve through the supernova phase.  
Such pulsars are expected to be produced and should have been
detected through existing surveys but they are not present.  
Given the low overall fraction of magnetars among field pulsars \citep[$\lsim 10^{-2}$; ][]{2014MNRAS.440L..86C},
identification of the first Galactic Center pulsar as a magnetar is surprising.  
Note
that while the X-ray outburst made quick radio discovery of \sgr\ possible, the many radio pulsar searches
over the past decades would have readily found the magnetar as well as the more stably emitting ordinary pulsars.
If massive stars in binary systems are common in the Galactic Center cluster and efficiently form magnetars,
however, the association of \sgr\ with the local stellar cluster may resolve this anomaly.  
This resolution requires that we are at an early stage of high-mass star formation following a period
of relative inactivity comparable to the characteristic pulsar lifetime ($\sim 10$ Myr) in which few pulsars were formed.  In this scenario,
the young pulsar population of the Galactic Center resembles that of a star-forming cluster similar to Wd 1, which has no
radio pulsars.  As the cluster ages, lower mass stars will go through a supernova phase and produce ordinary
pulsars (i.e., non-magnetars).

The star formation history of the Galactic Center cluster over the age of the Galaxy has been extensively studied 
\citep[e.g.,][]{2003ApJ...597..323B,2011ApJ...741..108P}.  Evolved giant branch stars appear to reveal
a significant rise in the star formation rate over the past few hundred million years, reaching the current peak
of $\sim 10^{-2}\ M_\odot$ yr$^{-1}$.  The star formation rate 100 Myr in the past is an order of magnitude
lower than the current rate.  The granularity of star-formation rate determination with this method is fairly coarse, however,
with characteristic widths of approximately $100$ Myr.  On the other hand, study of 
high mass stars indicates a burst of star formation no earlier than 6 Myr in the past \citep{2013ApJ...764..155L},
i.e., indicating variability in the star formation rate on time scales comparable to the pulsar lifetime.

Magnetar formation is poorly understood and based on a relatively small sample of objects.
For instance, evidence has been presented that some ordinary pulsars may evolve into magnetars
\citep{2011ApJ...741L..13E}.
Considering also the complex star-formation history of the Galactic Center,  we cannot present a definitive
account of the origin of \sgr\ and pulsars in the vicinity of Sgr A*.
Nevertheless, the anomaly of the first Galactic Center pulsar appearing as a magnetar and the apparently insufficient 
strength of the scattering screen to obscure ordinary pulsars is plausibly resolved through
a scenario in which magnetars are the earliest pulsars formed in the young Galactic Center cluster.

Millisecond pulsars (MSPs) from previous generations of star formation, on the other hand,
 should be present and would still have escaped
detection in existing surveys.  Even with the reduced temporal broadening estimate based on a greater
distance to the scattering screen,  detection of MSPs requires high frequency, high
sensitivity observations.  MSPs, therefore, provide the potential for the
characterization of general relativistic effects associated with Sgr A*.
Efforts are underway to search for MSPs at frequencies above 10 GHz with the 
VLA, GBT, and Effelsberg
\citep{2013IAUS..291...57S,2013IAUS..291..382E}; an SKA-sized array built with high frequency receivers or ALMA \citep{2013arXiv1309.3519F} 
may have the capability to detect MSPs if current efforts are unsuccesful.

\acknowledgements
The National Radio Astronomy Observatory is a facility of the National Science Foundation operated under cooperative agreement by Associated Universities, Inc. 


\end{document}